\documentclass[12pt]{article}

\usepackage[titletoc,title]{appendix}
\usepackage[latin1]{inputenc}
\usepackage{amsmath}
\usepackage{amssymb}
\usepackage{graphicx}
\usepackage[margin=1.25in]{geometry}
\usepackage[bottom]{footmisc}
\usepackage{indentfirst}
\usepackage{endnotes}
\usepackage{mathabx}
\usepackage{enumerate}
\usepackage{rotating}
\usepackage{multirow}
\usepackage{booktabs,calc}
\usepackage{threeparttable}
\usepackage{threeparttable}
\usepackage{float}
\usepackage{url}
\usepackage{epsfig}
\usepackage{caption,subcaption}
\usepackage{float}
\usepackage{url}
\usepackage{amsmath,amsfonts}
\usepackage{amsthm}
\usepackage{mathrsfs}
\usepackage{kbordermatrix}

\usepackage[onehalfspacing]{setspace}
\usepackage{apptools}

\newcommand{\suchthat}{\;\ifnum\currentgrouptype=16 \middle\fi|\;}

\newtheorem{theorem}{Theorem}
\newtheorem{prop}{Proposition}

\newtheorem{cor}{Corollary}

\newtheorem{example}{Example}
\theoremstyle{remark}

\usepackage{chngcntr}
\usepackage{graphicx}
\usepackage{amsmath}
\usepackage{amssymb}
\usepackage{parskip}
\usepackage{sgame}
\usepackage{color}
\usepackage{tikz}
\usetikzlibrary{trees,calc}

\DeclareMathOperator*{\argmax}{\arg\!\max}

\newcommand{\eps}{\varepsilon}
\newcommand{\supp}{\operatorname{supp}}

\usepackage{natbib}

\begin{document}

\title{\large{Non-rationalizable Individuals, Stochastic Rationalizability, and Sampling}\thanks{We thank Victor Aguiar, Roy Allen, Paul J. Healy, Nail Kashaev, Matthew Kovach, and participants at Midwest Theory Conference and OSU Theory/Experimental reading group for helpful comments.}}

\author{
    \small{Changkuk Im}  \\
    \small{Department of Economics} \\
    \small{The Ohio State University} \\
    \small{im.95@osu.edu}\\
    \and
    \small{John Rehbeck}  \\
    \small{Department of Economics} \\
    \small{The Ohio State University} \\
    \small{rehbeck.7@osu.edu}\\
}
\date{\small{ \today }}

\maketitle
\begin{abstract}
Experimental work regularly finds that individual choices are not deterministically rationalized by well-defined preferences. Nonetheless, recent work shows that data collected from many individuals can be stochastically rationalized by a distribution of well-defined preferences. We study the relationship between deterministic and stochastic rationalizability. We show that a population can be stochastically rationalized even when half of the individuals in the population cannot be deterministically rationalized. We also find the ability to detect individuals who are not deterministically rationalized from population level data can decrease as the number of observations increases.\\       

\noindent\emph{JEL Classification Numbers:} C00, D01, D11 \\
\emph{Keywords:} Stochastic rationalizability, revealed preference, demand types, sampling
\end{abstract}

\newpage
\setlength\parindent{24pt}

\section{Introduction}
Experimental and empirical studies regularly find that individuals make choices that cannot be rationalized by well-defined preferences.\footnote{We take a preference ordering as a complete and transitive weak preference order. Individual choices are rationalizable by a preference ordering when their choices satisfy the revealed preference conditions of \cite{richter1966revealed},  \cite{afriat1967construction}, and \cite{varian1982nonparametric}. These conditions are empirically violated in numerous domains such as household consumption \citep{echenique2011money, cherchye2013nash, demuynck2018revealed}, risk and uncertainty \citep{choi2007consistency, choi2014more, carvalho2016poverty, carvalho2019complexity, feldman2020revealing}, altruistic allocation problems \citep{andreoni2002giving, fisman2007individual}, and so on.} These findings suggest that datasets with many individuals are likely to contain some who make choices that are not rationalizable by any preference order. However, recent studies show that large cross-sectional datasets can be rationalized by a distribution of individuals with well-defined preferences \citep{kitamura2018nonparametric, deb2019revealed}. Although these two observations seem to contradict one another, this paper theoretically shows how this is possible by examining the relationship between deterministic and stochastic rationalizability. Throughout the paper, the concept of ``deterministic'' rationalizability applies to choices by an \emph{individual} while the concept of  ``stochastic'' rationalizability applies to \emph{aggregate} choices from a population or a sampled group.\footnote{Of course, the concept of stochastic rationalizability can be applied to an individual's choices. For the case of an individual, one could think of the individual having a probability of behaving like a given demand type.}

Throughout the paper, we focus on datasets with two consumption goods and two observations for simplicity. First, we show that even when there are individuals who are not deterministically rationalized in a population, the population dataset can still be stochastically rationalized. We say that this is a \emph{false acceptance of stochastic rationalizability}. We find that false acceptances can occur even when half of the population make choices that cannot be deterministically rationalized. Practically, this means that only populations with a majority of individuals who are not deterministically rationalizable are guaranteed to be detected as not stochastically rationalizable. Thus, one should be cautious when drawing conclusions from stochastic rationalizations since a large fraction of choices may be generated by individuals who are not rationalized by any preferences. 

Next, we take this insight to examine how distinct sampling schemes interact with the relationship between deterministic and stochastic rationalizability. For cross-sectional sampling, we find that even when all individuals cannot be deterministically rationalized, there exist random samples of the population that are stochastically rationalized. We also find that there are cross-sectional samples that can lead to a \emph{false rejection of stochastic rationalizability} where a researcher erroneously rejects that the sample dataset is stochastically rationalizable even when the population consists of only deterministically rationalizable individuals. For panel sampling, we find that if a researcher ignores the panel structure and does not examine rationality for each individual, then false acceptance persists even when almost all individuals in the population cannot be deterministically rationalized. However, panel sampling eliminates false rejections.\footnote{\cite{aguiar2018stochastic} have a dedicated method for panel sampling that uses information from individual choices and is not subject to this critique.} 

In addition, we provide a method to account for the ``power" of stochastic rationalizability by using a multinomial sampling scheme of estimated demand types. This is an analogue to comparing deterministic rationalizability to the ``power" of a random sample to reject the model following \cite{bronars1987power}. One interesting finding from our examples with two baseline population distributions is that it can be difficult to detect deterministically non-rationalizable individuals when checking a stochastic rationalization as the sample size increases. That is, a researcher may lose substantial ``power'' of stochastic rationality even one has a large dataset.

Our findings in this paper contribute to the existing literature in several aspects. First, we clarify the relationship between deterministic rationality from \cite{afriat1967construction} and \cite{varian1982nonparametric} to stochastic rationality from \cite{mcfadden1990stochastic} and \cite{mcfadden2005revealed}. In this sense, the recent paper of \cite{ok2020indifference} is related since it studies the relationship between deterministic rationality and stochastic choices. However, they focus on the rationalization of stochastic choices by allowing incompleteness of deterministic preferences or experimentation. In contrast, we assume completeness and heterogeneity of deterministic preferences. Based on this setting, we investigate whether individuals whose choices are not deterministically rationalized can be part of a population that is stochastically rationalized. Hence, our findings are closely related to \cite{becker1962irrational} since we show that aggregate data of many individuals may be categorized as rational even when a fraction of the population cannot be rationalized. 

Second, our paper helps clarify that stochastic rationalizability of a population is an ``as if" model. For example, \cite{hoderlein2014revealed} show how to bound the fraction of individuals who violate the weak axiom of revealed preference using cross-section data. \cite{hoderlein2014revealed} summarize how their results relate to the stochastic choice literature and state that one can use methods from \cite{mcfadden1990stochastic} to bound the largest number of irrational types consistent with a stochastic rationalization. Thus, the results here were ``folk knowledge" which we hope to make clearer to a larger audience by supposing there is a ``true population" and relating it to the recovered rational demand types that result when finding stochastic rationalizations. This also helps clarify the seeming contradiction between experimental studies violating revealed preference conditions and population studies that find a distribution of rationalizable individuals.

The results are also important for other studies building on the random utility model when trying to make counterfactual or welfare comparisons. For example, \cite{deb2019revealed} make welfare comparisons based on the proportion of deterministically rationalizable demand types recovered from the aggregate choice dataset. Since our results suggest the possibility of false categorization for the proportion of demand types when looking for a stochastic rationalization, however, a researcher may incorrectly estimate the proportion of rational individuals and obtain erroneous counterfactual and welfare estimates. \cite{kashaev2020discerning} discuss similar issues when studying solution concepts in games. For example, they look at when it is possible that one can ensure data are generated by a solution concept and not other behavior.

The rest of this paper is organized as follows. Section~\ref{sec:def} reviews the definitions of deterministic and stochastic rationalizability based on the strong axiom of revealed preference. Since two budget sets and two consumption goods cases are sufficient to discuss our main research questions, we focus on this setting throughout the paper. Section~\ref{sec:population} provides an intuitive example and the main results without sampling error. Section \ref{sec:sample} extends the main results by considering cross-sectional and panel sampling schemes and suggests the power computation in the spirit of Bronar's measure from a multinomial sampling scheme. Section~\ref{sec:conclusion} provides our final remarks.

\section{Definitions}\label{sec:def}
\begin{figure}[H]
      \begin{subfigure}[b]{0.48\textwidth}
        \begin{tikzpicture}
         \coordinate (O) at (0,0);
         \draw[->] (0,0) -- (5,0) coordinate[label = {below:$x_1$}] (xmax);
        \draw[->] (0,0) -- (0,5) coordinate[label = {left:$x_2$}] (ymax);
    
        \draw (0,4) -- (2,0);
        \draw (0,2) -- (4,0);

        \draw[->] (1,2) -- (1.6,2.3) node[right]{$p^1$};
        \draw[->] (3,0.5) -- (3.3,1.1) node[right]{$p^2$};

        \draw (1.15,2.075) -- (1.225, 1.925) -- (1.055,1.85) ;
        \draw (3.075,0.65) -- (3.225, 0.575) -- (3.15,0.425) ;
    \end{tikzpicture}
    \caption{Two overlapping budget sets}\label{fig:main}
    \end{subfigure}
    \begin{subfigure}[b]{0.48\textwidth}
    \begin{tikzpicture}
      \coordinate (O) at (0,0);
      \draw[->] (0,0) -- (5,0) coordinate[label = {below:$x_1$}] (xmax);
      \draw[->] (0,0) -- (0,5) coordinate[label = {left:$x_2$}] (ymax);
    
    \draw (0,4) -- (2,0);
    \draw (0,2) -- (4,0);

    \draw (.5,4.2) node{$x^{1|1}$};
    \draw (2.2,.5) node{$x^{2|1}$};
    \draw (.5,2.2) node{$x^{1|2}$};
    \draw (4,.5) node{$x^{2|2}$};
    \draw (1.7,1.7) node{$x^{3|t}$};
    \filldraw (1.33,1.33) circle (2pt);

    \draw[->] (1,2) -- (1.6,2.3) node[right]{$p^1$};
    \draw[->] (3,0.5) -- (3.3,1.1) node[right]{$p^2$};

    \end{tikzpicture}
    \caption{Demand regions}\label{fig:types}
    \end{subfigure}
        \caption{Two consumption good and two budget environment}\label{fig:all}
    \raggedright
\end{figure}
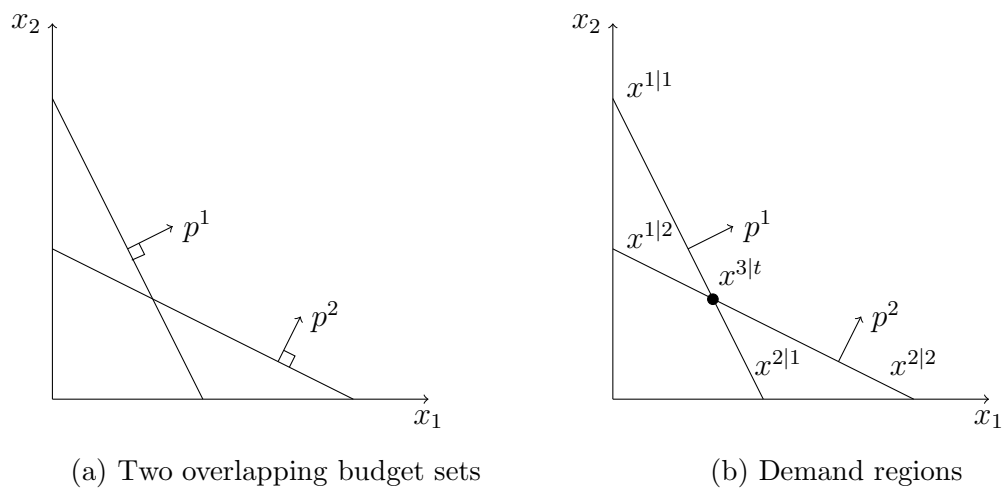

Here we define the standard consumer problem and the extension to random utility models. We study random utility models \citep{mcfadden1990stochastic, mcfadden2005revealed} for the standard consumer problem following \cite{hoderlein2015testing} and \cite{kitamura2018nonparametric}. It is enough to consider when there are two goods to discuss our main research question. Moreover, insights on the relationship between deterministic and stochastic rationalizability are most clearly seen when there are two budget sets as shown in Figure~\ref{fig:main}.\footnote{More generally the results here hold for any number of goods for two budgets that intersect. This occurs since larger numbers of goods produce (more-or-less) the same demand types that encode when each hyperplane is above or below the other.} Throughout the paper, we assume that budget lines overlap so that violations of rationality can be detected.\footnote{Formally, this means that for normalized prices from observation one ($p^1$) and observation two ($p^2$) that there exists a bundle $\hat{x} \in \mathbb{R}_+^2$ with $p^1 \cdot \hat{x} = p^2 \cdot \hat{x}$.}

\subsection{Deterministic Rationalizability}
A deterministic dataset of interest is given by $\mathcal{D}_{D}=\{(p^1,x^1),(p^2,x^2)\}$ where normalized prices in the $t$-th observation are given by $p^t \in \mathbb{R}_{++}^2$ and the observed consumption bundle is given by $x^t \in \mathbb{R}_+^2$. We define the normalized budget set by $B(p^t)=\{x \in \mathbb{R}_+^2 \mid p^t \cdot x \le 1\}$. We consider the standard consumer problem with a locally non-satiated utility function $u: \mathbb{R}_+^2 \rightarrow \mathbb{R}$ that yields a unique maximizer.\footnote{A utility function is defined as locally nonsatiated when for any $x \in \mathbb{R}_{+}^2$ and any $\varepsilon > 0$ there exists $y \in \mathbb{R}_+^2$ with $||y-x|| \le \eps$ such that $u(y) > u(x)$.} Thus, if observed choices are rationalized by utility maximization, then  
\[ x^t = \argmax_{x \in B(p^t)} \{u(x)\} .\]
If a deterministic dataset is rationalized in this way, we also say the dataset is \emph{deterministically} rationalized.

It is known that choices are consistent with a non-satiated utility function with singleton demand when they satisfy the strong axiom of revealed preferences \citep{houthakker1950revealed} which is a strengthening of the generalized axiom of revealed preference \citep{afriat1967construction, varian1982nonparametric}.\footnote{For two budget sets and two goods, we note the weak axiom of revealed preference is equivalent to the strong axiom of revealed preference.} The strong axiom of revealed preference for a dataset with two observations requires that for two distinct consumption bundles, if the $t$-th bundle costs less than the $s$-th bundle at the prices from the $s$-th observation, then the $s$-th bundle must be strictly more expensive than the $t$-th bundle at the prices from the $t$-th observation.

\begin{prop}\label{prop:deterministic}
The dataset $\mathcal{D}_D$ is rationalized by a locally non-satiated utility function with unique maximizers if and only if for all $s,t \in \{1,2\}$ with $s \neq t$   
\begin{align*}
    x^s \neq x^t \;\; \text{and} \;\; p^s x^t \le  p^s x^s  \quad \text{implies} \quad p^t x^t < p^t x^s.
\end{align*} 
\end{prop}

\subsection{Stochastic Rationalizability}
A stochastic demand system dataset is given by $\mathcal{D}_{S}=\{(p^1,\pi^1),(p^2,\pi^2)\}$ where normalized prices in the $t$-th observation are given by $p^t \in \mathbb{R}_{++}^2$ and $\pi^t$ is a distribution over consumption bundles where $\supp(\pi^t) \subseteq B(p^t)=\{x \in \mathbb{R}_+^2 \mid p^t \cdot x \le 1\}$. In other words, $\pi^{t}$ indicates a choice probability for each consumption bundle at observation $t$. For simplicity, we assume that each distribution $\pi^t$ is defined over a finite number of consumption bundles. The support of the distribution $\pi$ denoted by $\supp(\pi)$ is the set of points assigned positive probability by $\pi$. 

We now define a random utility model following \cite{kitamura2018nonparametric}. First, let $\pi^t(O)$ be the probability that a choice bundle is in the measurable set $O \subseteq \mathbb{R}_+^2$ at observation $t$. Let $\mathscr{U}$ be the space of strictly quasiconcave locally non-satiated utility functions $u : \mathbb{R}_+^{2} \to \mathbb{R}$. A stochastic demand system dataset $\mathcal{D}_{S}$ is rationalized by a  \emph{random utility model} (RUM) when there is a probability measure $\rho$ over the space of functions $\mathscr{U}$ such that, for all $t \in \{1,2\}$:
\begin{equation}\label{eq:rum}
\pi^t(O) \ = \ \rho\Big(\big\{u \in \mathscr{U} : \argmax\nolimits_{x \in B(p^t)} u(x) \in O \big\}\Big),
\end{equation}
for any measurable subset $O \subseteq \mathbb{R}_+^{2}$. We also say a stochastic demand system dataset is \emph{stochastically} rationalized if the dataset is rationalized in this way. The $\argmax$ set is a singleton since $\mathscr{U}$ consists of strictly quasiconcave functions. Thus, the probability of choosing a bundle in set $O$ is equal to the probability of drawing a utility function that has a maximizer in set $O$. 

Locally non-satiated utility functions generate choices on the budget line with probability one. Figure~\ref{fig:types} shows that demand can fall in one of three regions for each budget line. For example, $x^{1|1}$ is the region from the first observation that is above the second budget line. Moreover, $x^{3|t}$ is the consumption bundle where the budget lines intersect. Let $x^{r\mid t}$ are choices in the $r$-th region of the $t$-th budget set. 

\cite{hoderlein2015testing} show stochastic rationality by a random utility model only depends on choice probabilities in each of these regions. Let $\pi^{r|t}$ denote the probability of choosing in the $r$-th region for the $t$-th observation. For the two budget sets and two goods case, we encode choice probabilities in  $\pi = (\pi^{1|1},\pi^{2|1},\pi^{3|1},\pi^{1|2},\pi^{2|2},\pi^{3|2})$ where $\sum_{r \in \{1,2,3\}} \pi^{r|t} = 1$ for all $t \in \{1,2\}$.  

When there are two budgets and two goods, it is easy to relate deterministic choices with the random utility model. Here, we consider a space of demand types. For example, an individual may choose from $x^{1|1}$ in the first budget and $x^{3|2}$ in the second budget. The demand type associated with there choices is denoted by $\theta(1,3)$ where the first entry corresponds to the demand region from the first budget and the second entry corresponds to the demand region from the second budget. This generates nine different demand types recorded in Table~\ref{tab:demandTypes} where six types are deterministically rationalizable.

\begin{table}[H]
    \centering
    \begin{tabular}{c| c| c|c|c}
     & Demand Type & Budget 1 & Budget 2 & Deterministically Rationalized \\
     \hline
    Type 1  & $\theta(1,1)$ & $x^{1|1}$ & $x^{1|2}$ & Yes \\ 
    Type 2  & $\theta(1,2)$ & $x^{1|1}$ & $x^{2|2}$ & Yes \\
    Type 3  & $\theta(2,2)$ & $x^{2|1}$ & $x^{2|2}$ & Yes \\
    Type 4  & $\theta(1,3)$ & $x^{1|1}$ & $x^{3|2}$ & Yes \\
    Type 5  & $\theta(3,2)$ & $x^{3|1}$ & $x^{2|2}$ & Yes \\
    Type 6  & $\theta(3,3)$ & $x^{3|1}$ & $x^{3|2}$ & Yes \\
    Type 7  & $\theta(2,1)$ & $x^{2|1}$ & $x^{1|2}$ & No \\
    Type 8  & $\theta(2,3)$ & $x^{2|1}$ & $x^{3|2}$ & No \\
    Type 9  & $\theta(3,1)$ & $x^{3|1}$ & $x^{1|2}$ & No \\
    \end{tabular}
    \caption{Demand Types}
    \label{tab:demandTypes}
\end{table}

\cite{mcfadden1990stochastic}, \cite{mcfadden2005revealed}, \cite{hoderlein2015testing}, and \cite{kitamura2018nonparametric} characterize conditions for stochastic rationalizability by a random utility model. Stochastic rationality requires one to find a probability distribution over rationalizable demand types that sum to the observed choice probabilities in each region. The set of rationalizable types is denoted by $RT=\{ \theta(1,1), \theta(1,2), \theta(2,2), \theta(1,3),  \theta(3,2), \theta(3,3) \}$. The set of all types is denoted by $AT$. For a set $S$, let $\Delta(S)$ be a probability distribution indexed over elements of the set. Then a probability distribution over rationalizable demand types is given by 
\[ \mu \in \Delta(RT)=\left\{ \mu \in \mathbb{R}_+^{|RT|} \mid \sum_{(j,k) \; \text{s.t.} \atop \theta(j,k) \in RT} \mu(j,k)=1 \right\} \]
where $\mu(j,k)$ is the probability of type $\theta(j,k)$. A dataset $\mathcal{D}_S$ is stochastically rationalized if and only if there exists    
a measure of rational demand types $\mu \in \Delta(RT)$ such that 
\begin{equation}\label{matrix:system}
\kbordermatrix{
       & \theta(1,1) & \theta(1,2) &  \theta(2,2) &
       \theta(1,3) &
       \theta(3,2) & \theta(3,3) \\
       &  1 & 1 & 0 & 1 & 0 & 0 \\
       &  0 & 0 & 1 & 0 & 0 & 0 \\
       &  0 & 0 & 0 & 0 & 1 & 1 \\
       &  1 & 0 & 0 & 0 & 0 & 0 \\
       &  0 & 1 & 1 & 0 & 1 & 0 \\
       &  0 & 0 & 0 & 1 & 0 & 1 \\
  }   
  \begin{bmatrix} \mu (1,1) \\ \mu (1,2) \\ \mu (2,2) \\ \mu(1,3) \\ \mu(3,2) \\ \mu (3,3) \end{bmatrix}
  =
  \begin{bmatrix} \pi^{1|1} \\  \pi^{2|1} \\ \pi^{3|1} \\ \pi^{1|2} \\ \pi^{2|2} \\ \pi^{3|2} \end{bmatrix}.
\end{equation}
The linear program from  (\ref{matrix:system}) shows that the choice probability of the first region from the first observation, $\pi^{1|1}$, is equal to the sum of probabilities of demand types $\theta(1,1)$, $\theta(1,2)$, and $\theta(1,3)$. These are the types that choose in the first region of the first budget. The following proposition summarizes the characterization of \cite{hoderlein2015testing}.

\begin{prop}\label{prop:srational}
Consider a dataset $\mathcal{D}_{S}=\{(p^1,\pi^1),(p^2,\pi^2)\}$. The following statements are equivalent:
\begin{enumerate}[(i)]
    \item $\mathcal{D}_{S}$ is stochastically rationalized.
    \item For data from $\mathcal{D}_S$, there exists a measure of rational demand types that satisfies the system of (\ref{matrix:system}).
    \item $\mathcal{D}_{S}$ satisfies $\pi^{2|1} + \pi^{1|2} +  \pi^{3|1} +  \pi^{3|2} -\min \{ \pi^{3|1}, \pi^{3|2} \} \le 1$.
\end{enumerate}
\end{prop}

Proposition~\ref{prop:srational} shows the link between a distribution of rationalizable demand types and choice probabilities. It also shows how random utility directly restricts choice probabilities. If we assume that no individual chooses from the third region in both budgets, then  $\pi^{1|t} + \pi^{2|t} = 1 \text{ for all } t \in \{1,2\}$ and the third statement can be rewritten as $\pi^{1|1} \ge \pi^{1|2}$ or, equivalently, $\pi^{2|1} \le \pi^{2|2}$. 

\section{Relating Rationalizability Concepts}\label{sec:population}

This section examines the relationship between deterministic and stochastic rationalizability assuming no sampling error. First, we show an example where some individuals cannot be deterministically rationalized, but none-the-less the whole population is stochastically rationalized. This means that aggregate data can be stochastically rationalizable even when there are individuals who cannot be rationalized by any preference in a population. We call this a \emph{false acceptance of stochastic rationalizability}. Following the example, we analyze the general relationship.

\begin{example} \label{ex:main}
Let $p^1=(2,1)$ and $p^2=(1,2)$ be given normalized prices so budgets $B(p^1)$ and $B(p^2)$ overlap as in Figure~\ref{fig:types}.

Consider a population with two demand types where $90\%$ of the population is type $\theta(1,2)$ and $10\%$ is type $\theta(2,1)$. Type $\theta(1,2)$ individuals are deterministically rationalizable, but type $\theta(2,1)$ individuals are not deterministically rationalizable. This population generates the distribution of choices where $\pi=(\pi^{1|1},\pi^{2|1},\pi^{3|1},\pi^{1|2},\pi^{2|2},\pi^{3|2})=(\frac{9}{10},\frac{1}{10},0,\frac{1}{10},\frac{9}{10},0)$. This dataset is stochastically rationalized since it satisfies Proposition~\ref{prop:srational}. Indeed, a vector $\mu \in \Delta(RT)$ with $\mu(1,1)=\frac{1}{10}$, $\mu(1,2)=\frac{8}{10}$, $\mu(2,2)=\frac{1}{10}$, and $\mu(1,3) = \mu(3,2) = \mu(3,3) = 0$ solves the linear program in Equation (\ref{matrix:system}).

Relative to the true demand types that generate the population, the stochastic rationalization under-estimates the proportion of type $\theta(1,2)$, while over-estimating the proportion of types $\theta(1,1)$ and $\theta(2,2)$ in the population. 
\end{example}

Example~\ref{ex:main} shows that a stochastic demand system dataset can fail to refute stochastic rationalizability even when there are individuals who are not deterministically rationalized, i.e., type $\theta(2,1)$. Moreover, demand types of the stochastic rationalization can place positive probability on types that were not present in the population. 

We now present relevant analytical results of false acceptance for the population.
Let $\nu \in \Delta(AT) = \{ \nu \in \mathbb{R}^{|AT|}_+ \mid \sum_{j,k \in \{1,2,3\}} \nu(j,k) = 1 \}$ be a probability distribution over all demand types. For a population $\nu$, a sample dataset is equivalent to choices made by the given population without sampling error. For convenience, we say a population is stochastically rationalized whenever its dataset is stochastically rationalized. Using Proposition~\ref{prop:srational}, we obtain a characterization of stochastic rationality that depends on the probability of \emph{all} demand types in the population.

\begin{theorem}\label{thm:population_existence}
Consider a dataset $\mathcal{D}_S$ with budgets as in Figure~\ref{fig:types} and the researcher samples the entire population. A population distribution over demand types $\nu \in \Delta(AT)$ is stochastically rationalized if and only if 
\begin{align*}
    \sum_{k \in \{1,2,3\}} \nu(3,k) \le \sum_{j \in \{1,2,3\}} \nu(j,3) &\quad \text{implies} \quad \sum_{k \in \{1,2,3\}} \nu(2,k) \le \sum_{j \in \{1,2,3\}} \nu(j,2) \quad   \text{and}\\
    \sum_{k \in \{1,2,3\}} \nu(3,k) > \sum_{j \in \{1,2,3\}} \nu(j,3)
    &\quad \text{implies} \quad \sum_{j \in \{1,2,3\}} \nu(j,1) \le \sum_{k \in \{1,2,3\}} \nu(1,k).
\end{align*}

\end{theorem}

\begin{proof}[Proof of Theorem~\ref{thm:population_existence}]
By Proposition~\ref{prop:srational}, we know that a dataset is stochastically rationalized if and only if its observed choice probabilities satisfy \[ \pi^{2|1} + \pi^{1| 2} + \pi^{3|1}  + \pi^{3|2} -\min \{ \pi^{3|1}, \pi^{3|2} \} \le 1.\] Since a dataset has no sampling error, the probability of choosing in a region is the sum of probabilities for all relevant demand types including rationalizable and non-rationalizable ones. For example,  $\pi^{r|1} = \sum_{j \in \{1,2,3\}} \nu(r,j)$ and $\pi^{r|2} = \sum_{k \in \{1,2,3\}} \nu(k,r)$ for all $r \in \{1, 2, 3\}$. 

When $\sum_{k \in \{1,2,3\}} \nu(3,k) = \pi^{3|1} \le \pi^{3|2} = \sum_{j \in \{1,2,3\} } \nu(j,3)$, the stochastic rationalizability condition in Proposition~\ref{prop:srational} can be expressed as 
\begin{equation*}\begin{split}
    \pi^{2|1} + \pi^{1|2} + \pi^{3|2} 
    &= \sum_{k \in \{1,2,3\}} \nu(2,k) + \sum_{j \in \{1,2,3\}} \nu(j,1) + \sum_{j \in \{1,2,3\}} \nu(j,3) \\
    &= \sum_{k \in \{1,2,3\}} \nu(2,k) - \sum_{j \in \{1,2,3\}} \nu(j,2) + 1 \\
    &\le 1
\end{split}\end{equation*}
where the second equality follows since $\sum_{j,k \in \{1,2,3\}} \nu(j,k) = 1$. Re-arranging, we obtain
\begin{equation*}
    \sum_{k \in \{1,2,3\}} \nu(2,k) \le \sum_{j \in \{1,2,3\}} \nu(j,2).
\end{equation*}
Analogous algebra yields the second condition when $\sum_{k \in \{1,2,3\} } \nu(3,k) = \pi^{3|1} > \pi^{3|2} = \sum_{j \in \{1,2,3\}} \nu(j,3)$.
\end{proof}

Theorem~\ref{thm:population_existence} implies that we can directly determine the stochastic rationalizability of a population by observing its distribution of demand types. This complements the work of \cite{hoderlein2014revealed} since it gives a simple condition on the proportion of individual demand types that lead to stochastic rationalizations. Importantly, the conditions of Theorem~\ref{thm:population_existence} depend not only on rationalizable demand types, but \emph{all} demand types. The conditions critically depend on the probability of choices in the third region. If there is less probability in the third region of the first observation than the second observation, then the probability of choosing in the second region must be less in the first observation than the second observation for a stochastic rationalization to exist. The same logic applies to the complementary case. 

Demand types that choose in region three are knife-edge cases since the third region of each budget is a single point. Thus, it is important to understand how a population without these demand types behaves. We record results for this situation in the following corollary. 

\begin{cor}\label{cor:pop_no3}
Consider a dataset $\mathcal{D}_S$ with budgets as in Figure~\ref{fig:types} and the researcher samples the entire population. A population distribution over demand types $\nu \in \Delta(AT)$ that places no probability on demand types that choose in region three is stochastically rationalized if and only if $\nu(2,1) \le \nu(1,2)$.
\end{cor}

\begin{proof}[Proof of Corollary~\ref{cor:pop_no3}]
By Proposition~\ref{prop:srational} the demand types must satisfy $\pi^{2|1} + \pi^{1|2} \le 1$ which reduces to $\nu(2,1) + \nu(2,2) + \nu (1,1) + \nu(2,1) \le 1$ under restrictions on the demand types. Note that $\nu(1,2)=1-\nu (1,1) - \nu(2,2) - \nu(2,1)$ since these are the only demand types with positive probability. Thus, $\nu(2,1) \le \nu(1,2)$. 
\end{proof}

This condition can be quickly used in practice. For instance, in Example~\ref{ex:main}, we have $\nu(1,2) = \frac{9}{10}$ and $\nu(2,1) = \frac{1}{10}$ so that, by Corollary~\ref{cor:pop_no3}, we can immediately determine that the dataset is stochastically rationalized. There are some important properties of stochastic rationalizations that can be deduced using these conditions that we record in the following proposition.

\begin{prop}\label{prop:pop_accept_reject}
Suppose normalized prices give the demand regions in Figure~\ref{fig:types}.
\begin{enumerate}[(i)]
    \item All populations $\nu \in \Delta(AT)$ with $\nu(1,2) \ge \frac{1}{2}$ are stochastically rationalizable.
    \item All populations $\nu \in \Delta(AT)$ with $\nu(2,1) > \frac{1}{2}$ are not stochastically rationalizable.
    \item All populations with $\text{supp}(\nu) \in AT \setminus RT$ are not stochastically rationalizable.
\end{enumerate}
\end{prop}

\begin{proof}[Proof of Proposition~\ref{prop:pop_accept_reject}]
(i) Suppose that $\nu \in \Delta(AT)$ with $\nu(1,2) \ge \frac{1}{2}$. Simplifying the stochastic rationalizability conditions in Theorem~\ref{thm:population_existence} we require $\nu(2,1) + \nu(2,3) \le \nu(1,2) + \nu(3,2)$ or $\nu(2,1) + \nu(3,1) \le \nu(1,2) + \nu(1,3)$. Recall $\sum_{j,k \in \{1,2,3\} } \nu(j,k) = 1$ where $\nu(j,k) \ge 0$. It follows that $\nu(2,1) + \nu(2,3) \le 1 - \nu(1,2) \le \nu(1,2) \le \nu(1,2) + \nu(3,2)$ where $1-\nu(1,2)\le \nu(1,2)$ since $\nu(1,2) \ge \frac{1}{2}$. Similarly, $\nu(2,1) + \nu(3,1) \le 1 - \nu(1,2) \le \nu(1,2) \le \nu(1,2) + \nu(3,2)$. Hence, the given population is stochastically rationalized.

(ii) Suppose that $\nu \in \Delta(AT)$ with $\nu(2,1) > \frac{1}{2}$. Again, recall $\sum_{j,k \in \{1,2,3\}} \nu(j,k) = 1$ where $\nu(j,k) \ge 0$. The stochastic rationalizability conditions in Theorem~\ref{thm:population_existence} are not satisfied. To see this, note that $\nu(2,1) + \nu(2,3) \ge \nu(2,1) > 1 - \nu(2,1) \ge \nu(1,2) + \nu(3,2)$ where the strict inequality follows since $\nu(2,1)>\frac{1}{2}$. Similarly, for the second case  $\nu(2,1) + \nu(3,1) \ge \nu(2,1) > 1 - \nu(2,1) \ge \nu(1,2) + \nu(1,3)$. Thus, the population is not stochastically rationalized.

(iii) Since $\text{supp}(\nu) \in AT \setminus RT$ we have $\nu(1,2)=\nu(1,1)=\nu(2,2)=\nu(1,3)=\nu(3,2)=\nu(3,3)=0$. By Theorem~\ref{thm:population_existence}, stochastic rationalizability requires either $\nu(2,1)+\nu(2,3) \le 0$ or $\nu(2,1)+\nu(3,1) \le 0$ depending on the choice probabilities in region three. For the first case, if $\nu(2,3)>0$, then the population is not stochastically rationalizble. If $\nu(2,3)=0$, then $\nu(3,1) \le \nu(2,3)$ so $\nu(3,1)=0$ and $\nu(2,1)=1$ which is not stochastically rationalizable. The second case follows by analagous algebra.  
\end{proof}

The first part of Proposition~\ref{prop:pop_accept_reject} is a sufficient condition for a dataset to be stochastically rationalized. This condition is independent of other demand types in the population. Thus, a dataset where half of the population is not deterministically rationalizable can be stochastically rationalized when there are enough $\theta(1,2)$ types. The second part suggests that a rejection of stochastic rationalizablity is guaranteed when there is a large proportion of individuals who are not deterministically rationalized. The third part shows that if the population consists entirely of individuals who are not deterministically rationalized, then the dataset is not stochastically rationalizable. 

\section{Sampling Error and Rationalizability}\label{sec:sample}
This section investigates the interaction of deteministic and stochastic rationalizability when sampling error is involved by concerning different sampling schemes. First, we consider cross-section and panel sampling and see how sampling error can make the situation worse. Throughout the analysis, we assume that the panel structure is ignored and the rationality is not considered individually. Lastly, we suggest a formula computing ``power'' of stochastic rationalizability by concerning multinomial sampling scheme.

\subsection{Cross-section Sampling}
We interpret a cross-section sample of the data for each period to be a random sample from the population of individuals that is not necessarily related across periods except for being drawn from a common distribution. We describe this in more detail below. 

A random sample in the $t$-th period describes individuals sampled in the $t$-th observation. Let the random sample in period $t$ be denoted by $s^t \in S = \{ s^t \in \mathbb{R}_+^{|AT|} \mid s^t(j,k) \le \nu(j,k) \; \forall j,k \in \{1,2,3\} \}$ whose only restriction is the sample is less than or equal to the true proportion of individuals. If a researcher does not sample all individuals of a given type $\theta(j,k)$, then $s^t(j,k)<\nu(j,k)$. For example, $s^t(1,3)$ says that in the $t$-th period the researcher samples $\frac{s^t(1,3)}{\nu(1,3)}$ of all individuals who choose from region one when normalized prices are $p^1$ and from region three when normalized prices are $p^2$. Thus, cross-section sampling is defined by the samples in period one and two given respectively by $s^1,s^2 \in S$. The main feature of cross-section sampling is that the samples $s^1$ and $s^2$ do not need to be related in any particular way. 

We denote the observed stochastic datasets generated from a sample by $\hat{\pi}(s^1,s^2)$. In particular, for any $r$-th region, the observed choice probabilities are given by  $\hat{\pi}^{r\mid 1}(s^1,s^2)=\frac{\sum_{k \in \{1,2,3\} } s^1(r,k)}{\sum_{j,k \in \{1,2,3\} } s^1(j,k)}$ and $\hat{\pi}^{r\mid 2}(s^1,s^2)=\frac{\sum_{j \in \{1,2,3\} } s^2(j,r)}{\sum_{j,k \in \{1,2,3\} } s^2(j,k)}$. Here sampling in period one only affects the observed distribution of choices for observation one and sampling in period two only affects the observed distribution of choices for observation two. To check the stochastic rationalizability of the sample dataset, we can straightforwardly apply the results from Proposition~\ref{prop:srational}, i.e., $\hat{\pi}^{2|1} + \hat{\pi}^{1|2} + \hat{\pi}^{3|1}  + \hat{\pi}^{3|2} - \min \{ \hat{\pi}^{3|1}, \hat{\pi}^{3|2} \} \le 1$. Throughout the following results, we regularly drop dependence on the sample when discussing the sample dataset $\hat{\pi}$. 

The sample dataset can have little relation to the true percentage of demand types. The following proposition shows that there are cross-sectional samples that are stochastically rationalized even when all or an arbitrarily large proportion of individuals are not deterministically rationalized. These are examples of false acceptance of stochastic rationality generated by sampling error. Contrary to the case of perfect sampling, one can also reject stochastic rationality in the presence of cross-sectional sampling even when all individuals are deterministically rationalized. We call the rejection of stochastic rationality when all individuals are deterministically rationalizable \emph{false rejection of stochastic rationalizability}.

\begin{prop}\label{prop:cross}
Suppose normalized prices give the demand regions in Figure~\ref{fig:types}.
\begin{enumerate}[(i)]
    \item There exist populations $\nu \in \Delta(AT)$ with $\supp(\nu) \subseteq AT \setminus RT$ and cross-section random samples $s^1,s^2 \in S$ such that the dataset of prices and observed choice probabilities $\hat{\pi}$ is stochastically rationalized.
    \item For every $\eps \in (0,1)$, there exist populations $\nu \in \Delta(AT)$ with $\nu(2,1) = 1 - \eps$ and  cross-section random samples $s^1,s^2 \in S$ such that the dataset of prices and observed choice probabilities $\hat{\pi}$ is stochastically rationalized.
    \item There exist populations $\nu \in \Delta(AT)$ with $\supp(\nu) \subseteq RT$ and  cross-section random samples $s^1,s^2 \in S$ such that the dataset of prices and observed choice probabilities $\hat{\pi}$ is not stochastically rationalized.
\end{enumerate}
\end{prop}

\begin{proof}[Proof of Proposition~\ref{prop:cross}] 
(i) Suppose that $\nu \in \Delta(AT)$ with $\nu(3,1)=\frac{1}{2}$ and $\nu(2,3)=\frac{1}{2}$ so that no individual is deterministically rationalized. Consider the sample $s^1(3,1)=\frac{1}{2}$ and $s^1(2,3)=0$ and $s^2(3,1)=0$ and $s^2(2,3)=\frac{1}{2}$ generating the observed choice probabilities $\hat{\pi}^{3|1} = \hat{\pi}^{3|2} = 1$. Condition (iii) of Proposition~\ref{prop:srational} holds on the sampled distribution. Therefore, we conclude that the dataset is stochastically rationalized. In fact, the resulting stochastic demand system is deterministically rationalized.

(ii) Suppose that $\nu \in \Delta(AT)$ with $\nu(1,2) = \eps$ and $\nu(2,1) = 1 - \eps$. Consider the sample $s^1(1,2) = s^2(1,2) = \eps$ and $s^1(2,1) = s^2(2,1) =  0$. This generates the observed choice probabilities $\hat{\pi}^{1|1} = \hat{\pi}^{2|2} = 1$. Condition (iii) of Proposition~\ref{prop:srational} holds on the sampled distribution. Therefore, we conclude that the dataset is stochastically rationalized. In fact, the resulting stochastic demand system is deterministically rationalized.

(iii) Suppose that $\nu(1,1)=\frac{1}{2}$ and $\nu(2,2)=\frac{1}{2}$ so that all individuals are deterministically rationalized. The sample $s^1(1,1)=0$ and $s^1(2,2)=\frac{1}{2}$ and $s^2(1,1)=\frac{1}{2}$ and $s^2(2,2)=0$ is not stochastically rationalized since $\hat{\pi}^{2|1} + \hat{\pi}^{1|2} = 1 + 1 = 2 > 1$. In fact, the resulting stochastic demand system is not deterministically rationalized.
\end{proof}

Proposition~\ref{prop:cross} shows that rejecting or failing to reject stochastic rationalizability can greatly depend on the sampling scheme applied to the population. In particular, a population of all individuals who are not deterministically rationalized can generate stochastically rationalizable datasets. When restricted to a population where individuals do not choose in region three of either budget, almost all individuals can fail to be deterministically rational but produce a sample dataset that is stochastically rationalized. Lastly, even a population consists of only deterministically rationalizable individuals can fail to produce stochastically rationalizable datasets from a cross-section sample. 

An issue with cross-section sampling as shown through Proposition~\ref{prop:cross} is that one cannot guarantee that a representative group of individuals was sampled. This is an empirically relevant observation since some individuals are hard to reach which can result in sampling error. 

We later discuss how deterministic rationalizability, stochastic rationalizability, and multinomial sampling interact since one might assume demand types are selected into the sample independently. However, we show through simulation that even for large multinomial samples false acceptance of stochastic rationalizability can still regularly occur. To intuitively understand why this can occur, consider Example~\ref{ex:main}. Here if the population is sampled multinomially, then a researcher will obtain data that converges to the true proportion of individuals in the population. Nonetheless, the true proportion of the population still leads to a false acceptance of stochastic rationality. This final point does not rely on individuals making choices from the third region.  

\subsection{Panel Sampling}
Let $s^t$ be a random sample from the $t$-th observation as defined above. Panel sampling has the same individuals present in observation one and two. Thus, panel sampling is represented by $s^1=s^2$.\footnote{There are dedicated statistical methods to handle panel sampling studied in \cite{aguiar2018stochastic}.} Note that when $s^1=s^2$, the observed stochastic dataset $\hat{\pi}$ results from a convex combination of types in the support of the population. 

This section examines the dangers of not using the full structure of panel sampling. In particular, when a researcher has panel data they could look directly at deterministic rationality conditions for each individual which will lead to correct results. Alternatively, a researcher could look for a stochastic rationalization which throws away information on individual choices. Here we show that not using the panel structure when looking for a stochastic rationalization can lead to false acceptances of stochastic rationality.  However, panel sampling prevents false rejections of stochastic rationality.

\begin{prop}\label{prop:panel}
Suppose normalized prices give the demand regions in Figure~\ref{fig:types}.
\begin{enumerate}[(i)]
    \item For all populations $\nu \in \Delta(AT)$ with $\supp(\nu) \subseteq AT \setminus RT$ and panel random samples $s^1 = s^2 \in S$, the dataset of prices and observed choice probabilities $\hat{\pi}$ is not stochastically rationalized.
    \item For every $\eps \in (0,1)$, there exist populations $\nu \in \Delta(AT)$ with $\nu(2,1) = 1 - \eps$ and  panel random samples $s^1 = s^2 \in S$ such that the dataset of prices and observed choice probabilities $\hat{\pi}$ is stochastically rationalized.
    \item For all populations $\nu \in \Delta(AT)$ with $\supp(\nu) \subseteq RT$ and panel random samples $s^1 = s^2 \in S$, the dataset of prices and observed choice probabilities $\hat{\pi}$ is stochastically rationalized.
\end{enumerate}
\end{prop}

\begin{proof}[Proof of Proposition~\ref{prop:panel}] 
(i) Suppose that $\nu \in \Delta(AT)$ with $\supp(\nu) \subseteq AT \setminus RT = \{ \theta(2,1), \theta(2,3), \theta(3,1) \}$. Panel sampling gives a sample dataset that can be represented by a population with $\supp(\nu) \in AT \setminus RT$. By Proposition~\ref{prop:pop_accept_reject} (iii) the given population is not stochastically rationalized. 

(ii) The proof here is the same as the proof of (ii) in Proposition~\ref{prop:cross}.

(iii) Suppose that $\supp(\nu) \subseteq RT$ and $s^1=s^2$ is a random sample. For $(j,k)$ such that $\theta(j,k) \in RT$, let the probability over rational types be given by  $\mu(j,k)=\frac{ s^1(j,k) }{ \sum_{ (\tilde{j},\tilde{k}) \; \text{s.t.} \; \theta(\tilde{j},\tilde{k}) \in RT} s^{1}(\tilde{j},\tilde{k}) }$. This is a random utility model by definition and is stochastically rationalizable.
\end{proof}

Proposition~\ref{prop:panel} implies that panel sampling prevents the worst distortions of data. That is, when a population consists entirely of deterministically rationalizable or non-rationalizable individuals, then any dataset gives the correct stochastic rationalization result. However, for a population with mixed rationalizable and non-rationalizable demand types, a dataset still can be stochastically rationalized even when there is arbitrary small fraction of rationalizable individuals.

\subsection{Multinomial Sampling and Bronars Power}
In this subsection, we relate multinomial sampling scheme to the ``power" measure from \cite{bronars1987power} and suggest a formula computing the ``power'' of stochastic rationalizability. Note that the multinomial sampling process  has the convenient property that the sample average of observed types almost surely converges to the true population probabilities.

Let $\nu \in \Delta(AT)$ be the true distribution of all demand types in a population. For multinomial sampling with replacement, any demand type $\theta(j,k)$ with $j,k \in \{1,2,3\}$ is sampled with probability $\nu(j,k)$. For simplicity, we assume that the sample size of each observation is the same, denoted by $n \in \mathbb{N}$, and that samples for each observation are independent. We also only consider demand types that never choose the third region for any observation. Recall from Proposition~\ref{prop:srational} that, under the above assumptions, a sample dataset is stochastically rationalized if and only if $\hat{\pi}^{1|1} \geq \hat{\pi}^{1|2}$. Below we calculate the probability that a sample dataset is stochastically rationalized.

The computation is straight-forward, but tedious, so we provide details. Since the sample size for each observation are the same size, effectively we can turn the condition $\hat{\pi}^{1\mid 1} \ge \hat{\pi}^{1\mid2}$ into one that checks whether there are more choices from the sample in region one of the first budget than region one of the second budget. This realization produces a tractable formula to compute the probability of a stochastic rationalization.

To see how this works, suppose that for the second observation we see no sample choices in region one. Using the multinomial theorem for $n$ observations this occurs with probability $\binom{n}{0} \left( \nu(1,1) + \nu(2,1) \right)^{0} \left( \nu(1,2) + \nu(2,2) \right)^{n} $. Conditional on this sample, any sample choices for the first observation are stochastically rationalizable. Thus, at least $\binom{n}{0} \left( \nu(1,1) + \nu(2,1) \right)^{0} \left( \nu(1,2) + \nu(2,2) \right)^{n}$ proportion of samples are stochastically rationalized.

Next, suppose that the sample of observation two has one choice in the first region. The probability this occurs is $\binom{n}{1} \left( \nu(1,1) + \nu(2,1) \right)^{1} \left( \nu(1,2) + \nu(2,2) \right)^{n-1}$. For the sampled choices to be stochastically rationalizable, at least one choice from the first observation must be in region one. The probability this occurs is $\sum_{\ell=1}^{n} \binom{n}{\ell} \left( \nu(1,1) + \nu(1,2) \right)^{\ell} \left( \nu(2,1) + \nu(2,1) \right)^{n-\ell}$. Thus, multiplying these probabilities gives the probability a sample dataset is stochastically rationalized when one choice is in region one of the second observation.

We can iterate and sum the above procedure to find the probability a multinomial sample of size $n$ in both periods is stochastically rationalized. In particular, the probability of a size $n$ multinomial sample being stochastically rationalized is 
\begin{equation}\label{eq:pass}
    \sum_{i=0}^{n} \left( p^{1|2} \right)^{i} \left( p^{2|2} \right)^{n-i} \binom{n}{i} \left[ \sum_{\ell=i}^{n} \left( p^{1|1} \right)^{\ell} \left( p^{2|1} \right)^{n-\ell} \binom{n}{\ell}  \right]
\end{equation}
where $p^{j|1} = \nu(j,1) + \nu(j,2)$ and $p^{j|2} = \nu(1,j) + \nu(2,j)$ for all $j \in \{1,2\}$. The term in the brackets is the probability the sample from observation one has more choices in region one than the sample from observation two. If $\nu(2,1) > 0$, then (\ref{eq:pass}) indicates the probability of false acceptance of stochastic rationalizability. If $\nu(2,1) = 0$, then one less (\ref{eq:pass}) is the probability of false rejections of stochastic rationalizability.

The above calculations will allow us to generate information related to the ``power" of stochastic rationalizability in a sense closely related to \cite{bronars1987power}. In this paper, we interpret the ``power" as the probability a dataset is not stochastically rationalized when there are some individuals in the population who are not deterministically rationalizable, i.e., $\nu(2,1) > 0$. 

We explicitly calculate the power for two distinct baseline population probabilities. We assume $p^1 = (2,1)$ and $p^2 = (1,2)$. The populations we consider are:
\begin{enumerate}
    \item Uniform Distribution: We assume a uniform distribution over demand types, i.e., $\nu(1,1) = \nu(1,2) = \nu(2,2) = \nu(2,1) = \frac{1}{4}$.
    \item Proportional Choices: We assume $\nu(1,1) = \nu(2,2) = \frac{2}{9}$, $\nu(1,2) = \frac{4}{9}$, and $\nu(2,1) = \frac{1}{9}$. This is related to random behavior discussed by \cite{becker1962irrational} since the distribution is proportional to size of the budget regions.\footnote{To see this, note that for prices $p^1 = (2,1)$ and $p^2 = (1,2)$ the intersection of budget lines gives regions where the proportional size of regions are $\frac{1}{3}$ and $\frac{2}{3}$. If people choose uniformly over the budget line, then we have a benchmark of $\nu(2,1)=\frac{1}{9}$.}  
\end{enumerate}

These two methods closely follow the intuition of \cite{bronars1987power} and \cite{becker1962irrational}. Thus, we provide a method to correct for ``power" when evaluating stochastic rationalizability while accounting for the fact that individuals may not be deterministically rationalizable. The first distribution is more attractive since for moderately sized datasets it is computationally costly to compute all rationalizable demand types.\footnote{See \cite{kitamura2018nonparametric} and \cite{de2019nonparametric} for details.} We present the results of simulations for different sample sizes in Table~\ref{tab:sim}.

\begin{table}[H]
    \centering
    \begin{tabular}{c|ccccc}
     & \multicolumn{4}{c}{Sample Size} \\
      & 10 & 50 & 100 & 500 & 1,000 \\
     \hline
    Uniform Sampling  & 0.5881 & 0.5398 & 0.5282 & 0.5126 & 0.5089 \\
    Proportional Sampling  & 0.9624 & 0.9998 & 1.0000 & 1.0000 & 1.0000 
    \end{tabular}
    \caption{Probability a multinomial sample is stochastically rationalized according to two benchmark populations}
    \label{tab:sim}
\end{table}

Table~\ref{tab:sim} shows that the ability to detect when there are individuals who are not deterministically rationalized from stochastic choice data is low. In particular from the uniform sample simulations, we see that even when one fourth of the population is not deterministically rationalized, the population dataset is stochastically rationalizable over 50\% of the time and this does not improve much with large samples. That this is around 50\% likely results from a uniform sample being on the boundary of the condition from  Theorem~\ref{thm:population_existence}, i.e., $\nu(2,1) = \nu(1,2)$. 

The results are worse for the proportional sampling. Even though there is a substantial fraction of individuals who are not deterministically rationalizable $\left(\frac{1}{9}\right)$, it is almost impossible to detect this group of people. Moreover, the ability to detect this group of individuals worsens as the sample size increases. The reason this occurs is exactly because Theorem~\ref{thm:population_existence} holds on the population with strict inequality, i.e., $\nu(2,1) < \nu(1,2)$. Thus, as the sample grows larger, it becomes harder to detect individuals who are not rational.\footnote{In a statistical sense, this follows since the population distributions of the proportional choices are on the interior of the set of stochastically rationalizable datasets. Thus, the power to detect any violations of individual rationality vanishes as sample size grows.}

\section{Conclusion}\label{sec:conclusion}
This paper shows that it is difficult to detect violations of stochastic rationalizability even when there are large fractions of the population who are not deterministically rationalized. Thus, while stochastic choice models and non-parametric methods have risen in popularity, the old problems of aggregate behavior not representing individual behavior as mentioned in \cite{becker1962irrational} still re-appear for these methods and need to be addressed. These issues are especially important for counterfactual and welfare predictions since one will classify the demand types of individuals in a population incorrectly. We discussed how deterministic and stochastic rationality interact with random sampling and gave two baseline distributions to examine ``power" in the spirit of \cite{bronars1987power}.

There are several ways one could go from here. One could try to ``prune" data that is better excluded when making predictions. Alternatively, one could try to obtain counterfactual and welfare bounds that adapt at the aggregate level to individual errors. Some work in this direction is covered in \cite{allen2020satisficing} and \cite{allen2020counterfactual}. A third approach  might be to gather individual and aggregate data, examine predictions resulting from a common model for each dataset, and examine which method better predicts. This last approach suggests that while there might be many stories that fit with a model, there might be limited cases or levels of aggregation where the model is appropriate.

\newpage

\bibliographystyle{chicago}
\bibliography{ref}

\begin{thebibliography}{}

\bibitem[\protect\citeauthoryear{Afriat}{Afriat}{1967}]{afriat1967construction}
Afriat, S.~N. (1967).
\newblock The construction of utility functions from expenditure data.
\newblock {\em International economic review\/}~{\em 8\/}(1), 67--77.

\bibitem[\protect\citeauthoryear{Aguiar and Kashaev}{Aguiar and
  Kashaev}{2018}]{aguiar2018stochastic}
Aguiar, V.~H. and N.~Kashaev (2018).
\newblock Stochastic revealed preferences with measurement error.
\newblock {\em arXiv preprint arXiv:1810.05287\/}.

\bibitem[\protect\citeauthoryear{Allen and Rehbeck}{Allen and
  Rehbeck}{2020a}]{allen2020counterfactual}
Allen, R. and J.~Rehbeck (2020a).
\newblock Counterfactual and welfare analysis with an approximate model.
\newblock {\em arXiv preprint arXiv:2009.03379\/}.

\bibitem[\protect\citeauthoryear{Allen and Rehbeck}{Allen and
  Rehbeck}{2020b}]{allen2020satisficing}
Allen, R. and J.~Rehbeck (2020b).
\newblock Satisficing, aggregation, and quasilinear utility.
\newblock {\em Available at SSRN 3180302\/}.

\bibitem[\protect\citeauthoryear{Andreoni and Miller}{Andreoni and
  Miller}{2002}]{andreoni2002giving}
Andreoni, J. and J.~Miller (2002).
\newblock Giving according to garp: An experimental test of the consistency of
  preferences for altruism.
\newblock {\em Econometrica\/}~{\em 70\/}(2), 737--753.

\bibitem[\protect\citeauthoryear{Becker}{Becker}{1962}]{becker1962irrational}
Becker, G.~S. (1962).
\newblock Irrational behavior and economic theory.
\newblock {\em Journal of Political Economy\/}~{\em 70\/}(1), 1--13.

\bibitem[\protect\citeauthoryear{Bronars}{Bronars}{1987}]{bronars1987power}
Bronars, S.~G. (1987).
\newblock The power of nonparametric tests of preference maximization.
\newblock {\em Econometrica: Journal of the Econometric Society\/}, 693--698.

\bibitem[\protect\citeauthoryear{Carvalho and Silverman}{Carvalho and
  Silverman}{2019}]{carvalho2019complexity}
Carvalho, L. and D.~Silverman (2019).
\newblock Complexity and sophistication.
\newblock {\em NBER Working Paper\/}~(w26036).

\bibitem[\protect\citeauthoryear{Carvalho, Meier, and Wang}{Carvalho
  et~al.}{2016}]{carvalho2016poverty}
Carvalho, L.~S., S.~Meier, and S.~W. Wang (2016).
\newblock Poverty and economic decision-making: Evidence from changes in
  financial resources at payday.
\newblock {\em American Economic Review\/}~{\em 106\/}(2), 260--284.

\bibitem[\protect\citeauthoryear{Cherchye, Demuynck, and De~Rock}{Cherchye
  et~al.}{2013}]{cherchye2013nash}
Cherchye, L., T.~Demuynck, and B.~De~Rock (2013).
\newblock Nash-bargained consumption decisions: A revealed preference analysis.
\newblock {\em The Economic Journal\/}~{\em 123\/}(567), 195--235.

\bibitem[\protect\citeauthoryear{Choi, Fisman, Gale, and Kariv}{Choi
  et~al.}{2007}]{choi2007consistency}
Choi, S., R.~Fisman, D.~Gale, and S.~Kariv (2007).
\newblock Consistency and heterogeneity of individual behavior under
  uncertainty.
\newblock {\em American Economic Review\/}~{\em 97\/}(5), 1921--1938.

\bibitem[\protect\citeauthoryear{Choi, Kariv, M{\"u}ller, and Silverman}{Choi
  et~al.}{2014}]{choi2014more}
Choi, S., S.~Kariv, W.~M{\"u}ller, and D.~Silverman (2014).
\newblock Who is (more) rational?
\newblock {\em American Economic Review\/}~{\em 104\/}(6), 1518--1550.

\bibitem[\protect\citeauthoryear{De~Rock, Cherchye, Smeulders, et~al.}{De~Rock
  et~al.}{2019}]{de2019nonparametric}
De~Rock, B., L.~Cherchye, B.~Smeulders, et~al. (2019).
\newblock Nonparametric analysis of random utility models: Computational tools
  for statistical testing.
\newblock Technical report, ULB--Universite Libre de Bruxelles.

\bibitem[\protect\citeauthoryear{Deb, Kitamura, Quah, and Stoye}{Deb
  et~al.}{2019}]{deb2019revealed}
Deb, R., Y.~Kitamura, J.~K.-H. Quah, and J.~Stoye (2019).
\newblock Revealed price preference: theory and empirical analysis.
\newblock {\em arXiv preprint arXiv:1801.02702\/}.

\bibitem[\protect\citeauthoryear{Demuynck and Seel}{Demuynck and
  Seel}{2018}]{demuynck2018revealed}
Demuynck, T. and C.~Seel (2018).
\newblock Revealed preference with limited consideration.
\newblock {\em American Economic Journal: Microeconomics\/}~{\em 10\/}(1),
  102--131.

\bibitem[\protect\citeauthoryear{Echenique, Lee, and Shum}{Echenique
  et~al.}{2011}]{echenique2011money}
Echenique, F., S.~Lee, and M.~Shum (2011).
\newblock The money pump as a measure of revealed preference violations.
\newblock {\em Journal of Political Economy\/}~{\em 119\/}(6), 1201--1223.

\bibitem[\protect\citeauthoryear{Feldman and Rehbeck}{Feldman and
  Rehbeck}{2020}]{feldman2020revealing}
Feldman, P. and J.~Rehbeck (2020).
\newblock Revealing a preference for mixing: An experimental study of risk.
\newblock Working Paper.

\bibitem[\protect\citeauthoryear{Fisman, Kariv, and Markovits}{Fisman
  et~al.}{2007}]{fisman2007individual}
Fisman, R., S.~Kariv, and D.~Markovits (2007).
\newblock Individual preferences for giving.
\newblock {\em American Economic Review\/}~{\em 97\/}(5), 1858--1876.

\bibitem[\protect\citeauthoryear{Hoderlein and Stoye}{Hoderlein and
  Stoye}{2014}]{hoderlein2014revealed}
Hoderlein, S. and J.~Stoye (2014).
\newblock Revealed preferences in a heterogeneous population.
\newblock {\em Review of Economics and Statistics\/}~{\em 96\/}(2), 197--213.

\bibitem[\protect\citeauthoryear{Hoderlein and Stoye}{Hoderlein and
  Stoye}{2015}]{hoderlein2015testing}
Hoderlein, S. and J.~Stoye (2015).
\newblock Testing stochastic rationality and predicting stochastic demand: the
  case of two goods.
\newblock {\em Economic Theory Bulletin\/}~{\em 3\/}(2), 313--328.

\bibitem[\protect\citeauthoryear{Houthakker}{Houthakker}{1950}]{houthakker1950revealed}
Houthakker, H.~S. (1950).
\newblock Revealed preference and the utility function.
\newblock {\em Economica\/}~{\em 17\/}(66), 159--174.

\bibitem[\protect\citeauthoryear{Kashaev and Salcedo}{Kashaev and
  Salcedo}{2020}]{kashaev2020discerning}
Kashaev, N. and B.~Salcedo (2020).
\newblock Discerning solution concepts for discrete games.
\newblock {\em Journal of Business \& Economic Statistics\/}, 1--14.

\bibitem[\protect\citeauthoryear{Kitamura and Stoye}{Kitamura and
  Stoye}{2018}]{kitamura2018nonparametric}
Kitamura, Y. and J.~Stoye (2018).
\newblock Nonparametric analysis of random utility models.
\newblock {\em Econometrica\/}~{\em 86\/}(6), 1883--1909.

\bibitem[\protect\citeauthoryear{McFadden and Richter}{McFadden and
  Richter}{1990}]{mcfadden1990stochastic}
McFadden, D. and M.~K. Richter (1990).
\newblock Stochastic rationality and revealed stochastic preference.
\newblock {\em Preferences, Uncertainty, and Optimality, Essays in Honor of Leo
  Hurwicz, Westview Press: Boulder, CO\/}, 161--186.

\bibitem[\protect\citeauthoryear{McFadden}{McFadden}{2005}]{mcfadden2005revealed}
McFadden, D.~L. (2005).
\newblock {Revealed stochastic preference: A synthesis}.
\newblock {\em Economic Theory\/}~{\em 26\/}(2), 245--264.

\bibitem[\protect\citeauthoryear{Ok and Tserenjigmid}{Ok and
  Tserenjigmid}{2020}]{ok2020indifference}
Ok, E.~A. and G.~Tserenjigmid (2020).
\newblock Indifference, indecisiveness, experimentation and stochastic choice.
\newblock Working Paper.

\bibitem[\protect\citeauthoryear{Richter}{Richter}{1966}]{richter1966revealed}
Richter, M.~K. (1966).
\newblock Revealed preference theory.
\newblock {\em Econometrica: Journal of the Econometric Society\/}, 635--645.

\bibitem[\protect\citeauthoryear{Varian}{Varian}{1982}]{varian1982nonparametric}
Varian, H.~R. (1982).
\newblock The nonparametric approach to demand analysis.
\newblock {\em Econometrica\/}, 945--973.

\end{thebibliography}

\end{document}